\documentclass{aa}
\newcommand{\be} {\begin{equation}}

\newcommand{\ee} {\end{equation}}
\newcommand{\etal}{{\it et al. }}
\newcommand{\src}{2E\,0053.2--7242}
\newcommand{\R}{{\em ROSAT}}
\newcommand{\E}{{\em Einstein}}
\newcommand{\BSAX}{{\em Beppo}SAX} 
\newcommand{\RXTE}{{\em R}XTE}
\newcommand{\bc}{\begin{center}}
\newcommand{\ec}{\end{center}}
\input psfig.tex
\begin{document}

\thesaurus{ (08.02.3; 08.14.1; 08.16.7; 11.13.1; 13.25.5)}

\title{\R\ archival observations of 1SAX J0054.9--7226 = 2E 0053.2--7242, a newly 
discovered X--ray pulsar in the SMC}

\author{G.L. Israel\inst{1, }\thanks{Affiliated to I.C.R.A.} \and S. 
Campana\inst{2}$^{, \star}$ \and G. Cusumano\inst{3} 
\and F. Frontera\inst{4} \and 
M. Orlandini\inst{4} 
\and A. Santangelo\inst{3} \and L. Stella\inst{1}$^{, \star}$}

\institute{Osservatorio Astronomico di Roma, Via Frascati 33, 
I--00040 Monteporzio Catone (Roma), Italy
\and
Osservatorio Astronomico di Brera, Via E. Bianchi 46, 
I--23807  Merate (Lecco), Italy
\and
Istituto di Fisica Cosmica e Applicazioni all'Informatica, C.N.R., 
Via La Malfa 153, I--90146 Palermo, Italy
\and
Istituto Tecnologie e Studio Radiazioni Extraterrestri, C.N.R., 
Via Gobetti 101, I--40129 Bologna, Italy
}

\date{Received 1 April 1998 / Accepted 9 April 1998}
\offprints{G.L. Israel: gianluca@coma.mporzio .astro.it}
\maketitle
\markboth{Israel \etal; \R\ observations of \src}
         {Israel \etal; \R\ observations of \src}

\begin{abstract}
We analysed 13 archival \R\ PSPC and HRI observations 
which included the position of a newly discovered 59\,s X--ray pulsar in the 
Small Magellanic Cloud, 1SAX J0054.9--7226 = \src. The source was detected 
three times between 1991 
and 1996 at a luminosity level ranging from $\sim$8$\times$10$^{34}$ - 4$\times$  
10$^{35}$ erg s$^{-1}$ (0.1--2.4 keV). 
Highly significant pulsations at 59.072\,s were detected during the 1991 
October 8--9 observation. The \R\ period, together 
with those measured by 
\RXTE\ and \BSAX\ yields a period derivative of \.P= -- 0.016 s yr$^{-1}$. 
A much more accurate source position (10$^{\prime\prime}$ uncertainty) was 
obtained through the \R\ HRI detection on 1996 April restricting to three  
m$_ V$ $>$ 15.5 stars the likely counterpart of 1SAX J0054.9--7226 = \src.
\end{abstract}


\section{Introduction}
On 1998 January 20 during a \RXTE\ observation in the direction of the 
Small Magellanic Cloud (SMC), a previously unknown X--ray source,  namely 
XTE J0055--724, was detected at a flux level (2--10 keV) of 
$\sim$ 6.0 $\times$ 10$^{-11}$ erg s$^{-1}$ cm$^{-2}$. The source showed 
pulsations at a period of $\sim$59\,s (Marshall \& Lochner, 1998a).  
A previous \RXTE\ observation of the same field performed on 1998 January 12 
failed to detect the source.

In response to these findings, simultaneous \BSAX\ and \RXTE\ observations 
of a region including the \RXTE\ error circle ($\sim$10$^{\prime}$ radius) of XTE 
J0055--724, were carried out on 1998 Jannuary 28. The results of these observations 
are reported elsewhere (Santangelo \etal 1998a; Marshall \etal 1998b). 
Thanks to the spatial capabilities of the imaging X--ray concentrators on board  
\BSAX, an improved position ($\sim$40$^{''}$ radius) was obtained for 
the pulsating source, named 1SAX J0054.9--7226 (Santangelo \etal 
1998a,b). The new \BSAX\ error circle contains only the previously 
classified \R\ and \E\ X--ray sources 1WGA J0054.9--7226 and \src, which 
are likely the 
same source. In the following we adopt the earliest source name, i.e. 
\E's. 
\src\ is a variable X--ray source in the SMC, which was already 
considered a candidate High Mass X--ray Binary by Wang \& Wu (1992; 
source \#35), Bruhweiler et al.  (1987; source \# 9) and by White \etal 
(1994; in the WGACAT), based on its high spectral hardness.

We report in this letter on the results from the analysis of the 
Position Sensitive Proportional Counter (PSPC) and High Resolution Imager (HRI) 
observations from the \R\ public archive.

\section{\R\ observations}
\begin{table*}[tbh]
\begin{center}
\caption{\R\ observations of \src\ }
\begin{tabular}{clrcccl}
\hline \noalign{\smallskip} 
  Pointing   & Instr.&Expos.&Start Time&Stop Time&Count rate$^{\dag}$& Notes$^{\ddag}$ \\
Number&&(s)& & &     (cts s$^{-1}$)  & \\\noalign{\smallskip} 
\hline\noalign{\smallskip}
 600195A00 & PSPC  &16978 &1991 Oct  08 03:10 & Oct 09  02:38  &  0.048$\pm$0.002 & T$_{\rm eff}$$=$11820s; [1]\\
 600196A00 & PSPC  & 1346 &1991 Oct  09 03:06 & Oct 10 04:47  &$<$ 0.080          & T$_{\rm eff}$$=$740s\\
 600196A01 & PSPC  &22223 &1992 Apr 15 15:30 & Apr 25 16:41  &$<$ 0.010          & T$_{\rm eff}$$=$10920s; [2]\\
 600195A01 & PSPC  & 9443 &1992 Apr 17 17:01 & Apr 27 16:28   & 0.008$\pm$0.001 & T$_{\rm eff}$$=$7202s  \\
 600452N00 & PSPC  &14207 &1993 Apr 10 11:54 & Apr 25 01:20  &$<$ 0.030&          T$_{\rm eff}$$=$6815s  \\
 600453N00 & PSPC  &17593 &1993 May  09 07:17&  May 12 20:14  &$<$ 0.005  &        T$_{\rm eff}$$=$7330s  \\
 500142N00 & PSPC &  4909 &1993 May 12 22:41 & May 13 04:11  &$<$ 0.015    &      T$_{\rm eff}$$=$2743s  \\
 600452A01 & PSPC & 16663& 1993 Oct  01 08:24 & Oct 14 16:54 & $<$ 0.014    &      T$_{\rm eff}$$=$8509s; [3]  \\
 500250N00 & PSPC & 20845 &1993 Oct 14 21:10 & Oct 29 08:45  &$<$ 0.014     &     T$_{\rm eff}$$=$10430s; [4]\\ \noalign{\smallskip} 
\hline \noalign{\smallskip}
 400237A01 & HRI   & 1167 &1993 Apr 17 09:33&  Apr 17 10:08  &$<$ 0.008 & \\
 400237N00 & HRI   & 1096 &1992 Oct 23 13:41 & Oct 24 08:45  &$<$ 0.010&\\
 400237A02 & HRI   & 1301 &1994 Apr 15 10:26 & Apr 15 11:05  &$<$ 0.008&\\
 600810N00 & HRI   & 6726 &1996 Apr 26 02:09 & Jun 10 14:45  &  0.004$\pm$0.001 & \\ \noalign{\smallskip} 
\hline\noalign{\smallskip}
\end{tabular} 
\end{center}
\noindent $^{\dag}$ Errors are at $1\,\sigma$ level and upper limits at 
$3\,\sigma$.\\
\noindent $^{\ddag}$ The effective times, T$_{\rm eff}$, are vignetted and 
exposure corrected.\\
\noindent [1] --- Pulsations at a period of 59.072\,s.\\
\noindent [2] --- Strong contamination by 1WGAJ0054.0--7226 and 1WGAJ0053.8--7226 due to 
the degrading PSF at large off--axis angles.\\
\noindent [3] --- Contamination by 1WGAJ0053.8--7226.\\
\noindent [4] --- Strong contamination by 1WGAJ0053.8--7226.\\ 
\end{table*}
The PSPC (0.1--2.4 keV) and HRI (0.1--2.4 keV) detectors on board \R\ 
observed the SMC field including \src\ several times. In the \R\ public 
archive there are 13 observations performed between 1991 October  
and 1996 April: 9 were carried out with the PSPC and 4 with the HRI. 
The PSPC images, spectra and light curves were accumulated and corrected 
for the effective exposure map. This is particularly relevant to minimize the 
effects of the wobble in the pointing direction on count rate measurements 
of sources near the edge of the field of view or the ribs of 
the detector window support structure. 
The vignetting correction was also taken into account and 
the effective exposure time $T_{\rm eff}$ obtained for each PSPC pointing.

A sliding cell detection algorithm was used in order to  
characterise the physical parameters, such as position, count rate (90\% 
confidence level), S/N ratio, etc., of 2E 0053.2--7242 when detected, and to 
obtain a $3\,\sigma$ count rate upper limit in case of non--detections. 
Table\,1 summarises the results of this analysis.

\src\  was only detected on three occasions: twice with the PSPC 
(1991 October 8--9 and 1992 April 17--27) and once with the HRI (1996 
April 26 -- June 10). 

\subsection{PSPC data}
The \R\ event list and spectrum of \src\ were extracted from a circle of 
$\sim$1$^{\prime}$.5 radius (corresponding to an encircled energy of 
$\sim$95\%) around the X--ray position. 
On 1991 October 8--9 (sequence number 600195A00)
the source count rate was 0.047 cts s$^{-1}$, while 
on 1992 April 17--27 (600195A01) it had decreased to about 0.01 cts s$^{-1}$.
Of the $\sim$ 600 and $\sim$ 150 photons contained in the extraction circle 
of each of the two pointings, we 
estimated that about 45 and 20 photons derive from the background 
around the source, respectively. 
\begin{figure*}[tbh]
\centerline{\psfig{figure=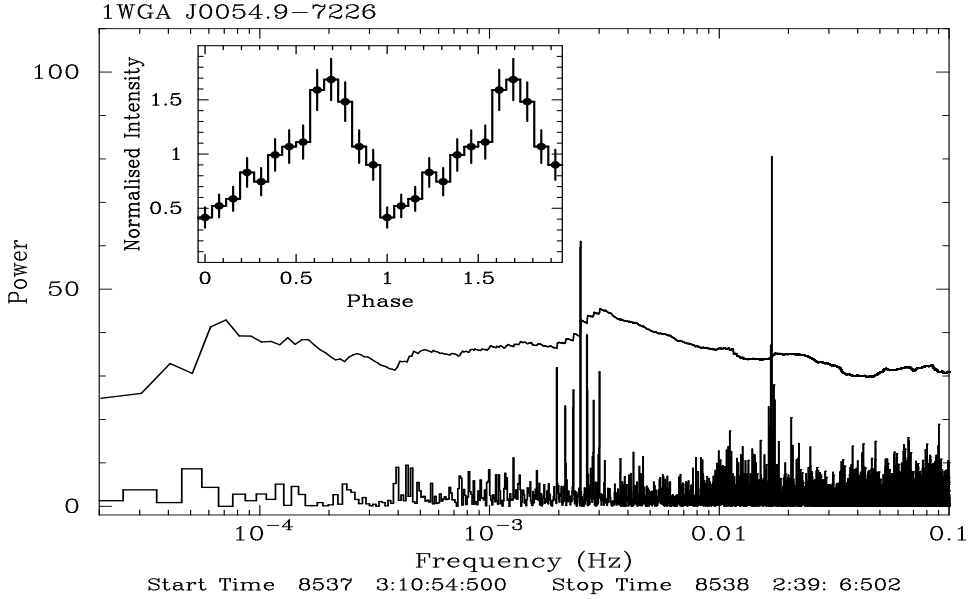,width=12cm,height=8.cm}}
\caption{Oct 1991 power spectrum of the 0.1--2.4 keV \R\ PSPC 
light curve of \src. The preliminary $3\,\sigma$ detection threshold 
is also shown. A folded light curve at the best period of 59.072\,s 
is shown as an insert. The peak structures around a frequency of $\sim$2.5 
$\times$ 10$^{-3}$ Hz are due to the wobble in the pointing direction}
\end{figure*}

The 1991 October observation is the only one that contains a sufficiently high 
number of  
photons to perform a detailed  periodicity search. The photons arrival times 
were corrected to the barycenter of the solar system and a 1\,s  binned light 
curve accumulated. The corresponding power spectrum, calculated over the entire 
observation duration ($\sim$1 day), is shown in Fig.\,1, together with the 
$3\,\sigma$ preliminary peak detection threshold described by Israel \& Stella 
(1996). The search was performed over a period interval around that detected 
by \BSAX\ and assuming a  maximum $|$\.P$|$ of $\sim$3 s yr$^{-1}$ (the highest 
ever observed from an X--ray pulsar: GX1+4). This translates into a search 
over an 
interval of $\sim$1100 Fourier frequencies centered on 0.017 Hz.
Significant peaks were detected around a frequency of $\sim$0.0169 Hz. 
These are unique to \src. The multiple peak  structure is due to the 
sidelobes arising from the satellite orbital occultation. 
The highest of these peaks has a significance of $7.4\,\sigma$ over the 
explored frequency interval, and corresponds to a period of 59.072 $\pm$ 0.003 s 
(90\% uncertainties are used through this letter). 
The modulation is energy independent in the PSPC band (within the statistical 
uncertainties). The shape is fairly asymmetric with a pulse fraction of $\sim $
40\% (Fig.\,1, inner panel). 
By using the period measured by \BSAX\ a period derivative of -- 0.016 s yr$^{-1}$ 
was obtained.

The PSPC Pulse Hight Analyser (PHA) rates were grouped so as to contain a 
minimum of 20 photons per energy bin.
The spectrum of the 1991 October observation was well fit with an 
absorbed power--law model (see Fig.\,2; upper curve). The best fit 
($\chi^2/$degree of freedom -- $dof$ = $21/24$) was obtained for a 
photon index of $\Gamma$ = 0.90 $\pm^{0.62}_{0.85}$ and a column density 
of N$_H$ = (1.3 $\pm^{1.6}_{0.8}$) $\times$ 10$^{21}$ cm$^{-2}$ 
(see Table\,2). Note that the Galactic hydrogen column in the direction 
of the SMC is $\sim$ 6 $\times$ 10$^{20}$ cm$^{-2}$. The 0.1--2.4 keV 
luminosity at the source is $L_X$ $\sim$ 4.2 $\times$ 10$^{35}$ erg s$^{-1}$ 
for an assumed distance of 65 kpc (Wang \& Wu 1992).
We note that any further spectral component added to the power--law model 
in order to fit the data excess around 1.2 keV did not significantly 
improve the fit.
For the 1992 April observation the number of photons (150) is too low to obtain 
an indipendent estimate of the spectral parameters. By keeping 
the photon index fixed to the best fit value of the 1991 October observation, 
an unabsorbed X--ray luminosity of $L_X$ $\sim$ 1.5 $\times$ 10$^{35}$ erg 
s$^{-1}$ was derived (see Table\,2 and Fig.\,2).

\begin{table}[bht]
\begin{center}
\caption{\R\ PSPC spectral results for \src\  }
\begin{tabular}{lll}
\hline  \noalign{\smallskip} \noalign{\smallskip} 
Parameter &  Oct 1991 & Apr 1992 \\ \noalign{\smallskip} 
\hline  \noalign{\smallskip} 
N$_H$ (10$^{22}$ cm$^{-2}$).......................... & 0.13$\pm_{0.08}^{0.16}$& 0.16$\pm_{0.16}^{0.55}$\\
$\Gamma$.................................................& 0.90$\pm^{0.62}_{0.85}$& 0.9\,(fixed) \\ \noalign{\smallskip} 
Count rate (10$^{-3}$ counts s$^{-1}$)..... &45.0$\pm$2.0& 8.0$\pm$1.0 \\
F$_X$ (10$^{-13}$ erg cm$^{-2}$ s$^{-1}$)........... &8.7 &2.4\\
$L_X$ (N$_H$=0; 10$^{35}$ erg s$^{-1}$)...........& 4.2& 1.5\\ \noalign{\smallskip} 
Reduced $\chi^2$/($dof$)....................... &0.96/(24)& ---\\ \noalign{\smallskip} 
\hline 
\end{tabular}
\end{center}
\noindent Note. --- the X--ray flux (absorbed) and the luminosity (unabsorbed) 
refer to the 0.1--2.4 keV energy band.
\end{table}
\begin{figure}[tbh]
\centerline{\psfig{figure=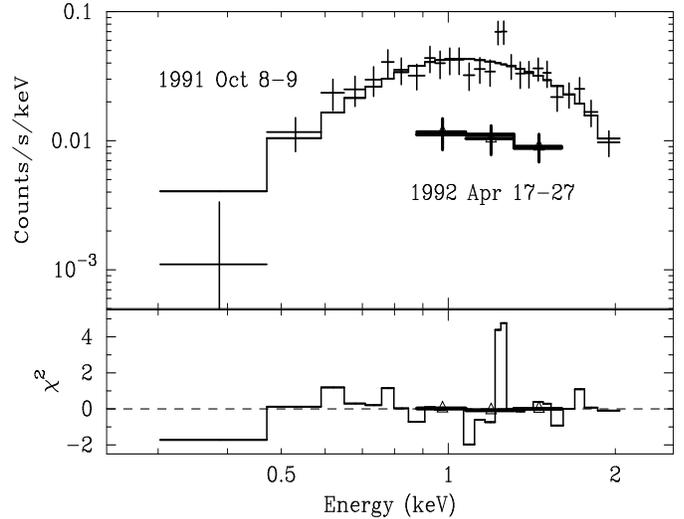,width=8cm,height=7.5cm}}
\caption{\R\ PSPC spectrum of \src\  during 1991 October and 1992 April. 
The best fit power--law is shown, together 
with the corresponding residuals. The 1992 Apr 17--27 data and residuals are 
marked with triangles}
\end{figure}

\subsection{HRI data}

The \R\ HRI observation during which \src\ was detected (sequence 600810; 
1996 April--June) provided a significantely improved source position (Israel 
1998).  This was determined to be  R.A. = 00$^h$ 54$^m$ 
56$^s$.5 and DEC = --72$^o$ 26$^{'}$ 47$^{''}$.3 (equinox 2000; error radius of 
1.1$^{\prime\prime}$ with a 90\% confidence level) by using both a sliding 
cell and a Wavelet transform--based detection algorithm (Lazzati et al. 1998; 
Campana et al. 1998).
However due to the unknown boresight correction the uncertainty radius 
increases up to 10$^{''}$.  

A 0.1--2.4 keV source flux level of $\sim$ 1.8$\times$ 
10$^{-13}$ erg cm$^{-2}$ s$^{-1}$ was determined assuming the best fit spectral 
parameters of the 1991 October 8--9. This translates into an unabsorbed 
luminosity of $L_X \sim 8.5$ $\times$ 10$^{34}$ (Israel 1998).

\section{\E\ observation}
\src\ was discovered during a survey of the SMC by means of the 
Imaging Proportional Counter (IPC) on board the \E\ spacecraft (Bruhweiler 
\etal, 1987; Wang \& Wu, 1992) and classified as an hard X--ray source based 
on its hardness ratio. It was observed at a count rate of 
$\sim$0.009$\pm$0.002 cts s $^{-1}$ corresponding to an X--ray luminosity 
of $\sim$1.5$\times$ 10$^{35}$ erg s$^{-1}$ (0.16--3.5 keV band). 

\section{Discussion}

\src\ was detected three times between 1991 and 1996 in the \R\ archival 
data. Highly significant pulsations, at a period of 59.072\,s were detected 
on 1991 October 8--9. These findings, together with the \BSAX\ results, yield 
a mean period derivative of $\sim$ -- 0.016\, s yr$^{-1}$ between 1991 and 
1998.  

\begin{figure}[tbh]
\centerline{\psfig{figure=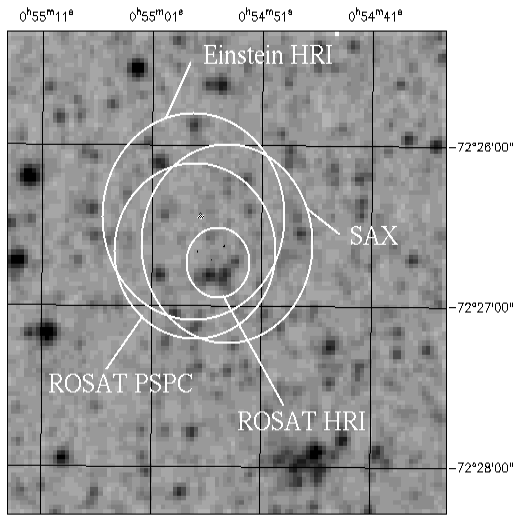,width=8.cm,height=8.cm}}
\caption{ESO plate including the position of \src. The X--ray error circles 
obtained from different instruments and satellites are shown}
\end{figure}

In one case a spectral analysis could be performed. The spectrum was found to 
be consistent with a relatively flat low absorbed power--law model that is 
typical of accreting X--ray pulsars in this energy range.

The 0.1--2.4 keV luminosity of \src\ as observed with \R\ 
ranges between $\sim$4.2$\times$10$^{35}$ erg 
s$^{-1}$ (1991 October 8--9) and $\sim$8.5$\times$10$^{34}$ erg s$^{-1}$ (1996 
April 26 -- June 10). Moreover \RXTE\ detected \src\ at a 
luminosity level of $\sim$3$\times$10$^{37}$ erg s$^{-1}$ in the 2--10 keV energy 
band. Extrapolating to the \R\ energy range the luminosity measured by \RXTE\ 
on 1998 January 20, a 0.1--2.4 keV luminosity of $\sim$2.5$\times$10$^{36}$ 
erg s$^{-1}$ is derived, implying a pronounced long--term variability of  
\src\ (a factor of $>$30). This indicates that the source is probably a transient 
X--ray pulsar in a  high--mass binary containing a Be star.

A 10$^{\prime\prime}$ accurate position was obtained thanks to a \R\ HRI 
observation during which the source was detected (1996 April; 
0.1--2.4 keV luminosity of $\sim$8.5$\times$10$^{34}$ erg s$^{-1}$). 
The \R\ HRI error circle of contains only three stars in the ESO plates with 
m$_V$ $>$ 15.5, the likely optical counterpart of \src (see Fig.\,3). 
Assuming a B--V = --0.2 and a 
distance modulus of 19 mag, these optical counterpart candidates are 
consistent with main sequence A9 -- B2 stars. We note that a similar 
spectral--type star (B1.5Ve; m$_V$ = 16) is the companion of the nearby 
X--ray source SMC X--2. Future optical follow--up observations 
of these candidates should determine the counterpart of \src\ 
and its probable Be star X--ray transient nature. The optical and/or infrared 
activity brightening of the counterpart will allow further X--ray triggers 
and studies.

\begin{acknowledgements} 
We would like to thanks D. Dal Fiume, A.N. Parmar and L. Piro for their constant 
help and scientific support.  
This research has made use of data obtained through the High Energy
Astrophysics Science Archive Research Center (HEASARC), provided by 
NASA's Goddard Space Flight Center. This work was par\-tial\-ly sup\-por\-ted 
thro\-ugh grants from the Agenzia Spaziale Italiana.
\end{acknowledgements}

\end{document}